# Breather solution of non-linear Klein-Gordon equation

D.V. Zav'yalov[1], V.I. Konchenkov[1,2], S.V. Kryuchkov[1,2]

[1] Volgograd State Technical University

[2] Volgograd State Socio Pedagogical University

**Abstract**

A technique for obtaining an approximate breather solution of the Klein-Gordon equation is presented. A breather solution of the equation describing the propagation of nonlinear waves in a graphene-based superlattice is investigated.

**Introduction**

The Klein-Gordon equation of the form

$$\frac{\partial^2 u}{\partial t^2} - \frac{\partial^2 u}{\partial x^2} + F(u) = 0, \qquad (1)$$

where $F(u)$ is an odd function with a period of $2\pi$, is found in various parts of theoretical and mathematical physics. One of the most important (and most common) special cases of this equation is the sine-Gordon equation (SG) [1 - 3]:

$$\frac{\partial^2 u}{\partial t^2} - \frac{\partial^2 u}{\partial x^2} + \sin u = 0. \qquad (2)$$

The SG equation is remarkable in that it has solutions in the form of solitary waves – solitons and breathers. The exact solution of equation (2) in the form of a traveling breather has the form [4, 5, 6]:

$$u_\omega(x,t) = 4\arctan\left(\frac{\sqrt{1-\omega^2}}{\omega}\frac{\cos\left(\gamma\omega t - \omega x\sqrt{\gamma^2-1}\right)}{\cosh\left(\gamma x\sqrt{1-\omega^2} - t\sqrt{1-\omega^2}\sqrt{\gamma^2-1}\right)}\right). \quad (3)$$

Here $\gamma = \dfrac{1}{\sqrt{1-\dfrac{v^2}{c^2}}}$, $v$ is the group velocity of the pulse propagation, $c$ is the speed of light.

The article [2] is devoted to the numerical study of kink and breather solutions of the sine-Gordon equation under the influence of a "force" determined by the Heaviside function $H(\xi)$:

$$\frac{\partial^2 u}{\partial t^2} - \frac{\partial^2 u}{\partial x^2} + \sin u = F(x,t), \quad (4)$$

$$F(x,t) = AH(t-x).$$

Boundary conditions of various types are considered in [2], and exact solutions of equation (2) are used as the initial condition for solving the perturbed equation (4).

In [3], the solutions of the two-dimensional sine-Gordon equation similar to the kink solutions of the one-dimensional equation (2) are investigated:

$$\frac{\partial^2 u}{\partial t^2} - \frac{\partial^2 u}{\partial x^2} - \sigma\frac{\partial^2 u}{\partial y^2} + \sin u = 0, \quad (5)$$

$\sigma = \pm 1$. Here the interaction of kink and antikink is considered, a system of equations is derived that allows determining the width and shape of both a solitary kink and soliton-like pulses interacting with each other. A numerical procedure for determining the center of the kink and a variational procedure for studying the dynamics of the shape of a single kink in the direction perpendicular to its propagation are discussed. The paper [7] is also devoted to the numerical study of stationary and traveling breather solutions of the two-dimensional sine-Gordon equation.

The article [8] is devoted to the study of real spatially periodic, centrally symmetric solutions of the Klein-Gordon equation of the form:

$$\frac{\partial^2 u}{\partial t^2} - \frac{\partial^2 u}{\partial x^2} - \frac{\partial^2 u}{\partial y^2} - \frac{\partial^2 u}{\partial z^2} + m^2 u - \Gamma(x) u^3 = 0 \tag{6}$$

The author of a paper [8] calls the obtained solutions as breathers by analogy with solutions of the sine-Gordon equation with similar properties. In [8], a reformulation of equation (6) was used as a system of coupled nonlinear Helmholtz equations under certain conditions in the field in the far zone.

A number of works are devoted to the study of exact and approximate breather solutions of various variants of the Klein-Gordon equation, for example, [9, 10]. The paper [9] considers the numerical solution of the discrete Klein-Gordon equation describing the oscillations of an infinite chain of particles associated with the nearest neighbors in a local potential [11]:

$$\frac{d^2 x_n}{dt^2} + V'(x_n) = \gamma (x_{n+1} + x_{n-1} - 2x_n). \tag{7}$$

In the long-wave approximation, equation (7) reduces to the Klein-Gordon equation (1).

In a paper [11], devoted to the study of vibrations in a one-dimensional chain of atoms, taking into account anharmonicity, a method of approximate solution of the Klein-Gordon equation in the limit of small amplitudes ($u \ll 1$) is proposed. In this case, equation (1) is transformed to the form

$$\frac{\partial^2 u}{\partial t^2} - \frac{\partial^2 u}{\partial x^2} + u - \beta u^3 = 0. \tag{8}$$

In [11, 12], a one-parameter localized periodic solution is sought in the form of

$$u = A(x)\cos(\omega t) + B(x)\cos(3\omega t) + ... \tag{9}$$

For the convergence of solution (9), it is necessary to put $|A| \gg |B| \gg ....$ In the case of small amplitudes, the largest contribution to the solution is provided by the first term, i.e. the solution is like a standing wave. Such solutions in the literature are sometimes called standing (stationary) breathers. For example, for the sine-Gordon equation, the solution [4] is known:

$$u_\omega(x,t) = 4\arctan\left(\frac{\sqrt{1-\omega^2}}{\omega} \frac{\sin(\omega t)}{\cosh\left(x\sqrt{1-\omega^2}\right)}\right), \tag{10}$$

This solution (10) in the limit $\omega \to 1$ can be approximately represented as $u = A(x)\cos(\omega t)$. The aim of this work is to obtain a solution in the form of a traveling breather of the nonlinear Klein-Gordon equation (8) using the methodology developed in [11, 12].

## 1. A technique for obtaining an approximate solution of the nonlinear Klein-Gordon equation in the form of a traveling small amplitude breather

Let's to consider obtaining an approximate analytical solution of the Klein-Gordon equation (1). As noted, in the case of $u \ll 1$ the equation (1) is transformed to the form (8). We will look for a solution to equation (8) in the form of a series (9) with uniformly decreasing coefficients before cosines. Substituting (9) into (8) and considering that $|A| \gg |B|$ (for the convergence of the series (9)), we obtain the system

$$\begin{cases} \dfrac{d^2 A}{dx^2} - (1-\omega^2)A = -\dfrac{3}{4}\beta A^3, \\ \dfrac{d^2 B}{dx^2} + (9\omega^2 - 1)B = -\dfrac{1}{4}\beta A^3. \end{cases} \tag{10}$$

Solving the system in the domain of limited in infinity localized solutions, it is possible to determine the functional form of a small-amplitude breather.

Let's try to construct, by analogy with the above, a method for finding traveling breathers of small amplitude, which are the solution of equation (8). In fact, the solution we are interested in is now a two-parameter delocalized periodic time solution. We will start from the known form of such a solution for the sine-Gordon equation (2), represented by the expression (3). We will look for a solution to equation (4) for a traveling breather of small amplitude in the form

$$u = A\left(\gamma x\sqrt{1-\omega^2} - t\sqrt{1-\omega^2}\sqrt{\gamma^2-1}\right)\cos\left(\gamma\omega t - \sqrt{\gamma^2-1}\omega x\right) +$$
$$B\left(\gamma x\sqrt{1-\omega^2} - t\sqrt{1-\omega^2}\sqrt{\gamma^2-1}\right)\cos\left(3\gamma\omega t - 3\sqrt{\gamma^2-1}\omega x\right) + \ldots \quad (11)$$

Now, following the previous method, after substituting (11) into equation (8) and some manipulations, we obtain a system for determining functions $A(\zeta)$, $B(\zeta)$, (hereinafter the notation is introduced $\zeta = \gamma x\sqrt{1-\omega^2} - t\sqrt{1-\omega^2}\sqrt{\gamma^2-1}$):

$$\begin{cases} (1-\omega^2)\dfrac{d^2A}{d\zeta^2} - (1-\omega^2)A = -\dfrac{3}{4}\beta A^3, \\ (1-\omega^2)\dfrac{d^2B}{d\zeta^2} + (9\omega^2-1)B = -\dfrac{1}{4}\beta A^3. \end{cases} \quad (12)$$

The solution of the first equation in (12) suitable for us by properties is the function

$$A(\zeta) = \left(\frac{8}{3\beta}(1-\omega^2)\right)^{1/2} \frac{1}{\cosh(\zeta)}. \quad (13)$$

For the sine-Gordon equation $\beta = 1/6$, and in the higher approximation we obtain the solution

$$u = 4\sqrt{1-\omega^2}\,\frac{\cos\left(\gamma\omega t - \sqrt{\gamma^2-1}\omega x\right)}{\cosh\left(\gamma x\sqrt{1-\omega^2} - t\sqrt{1-\omega^2}\sqrt{\gamma^2-1}\right)}. \quad (14)$$

It can be seen from (3) that, in fact, the condition $u \ll 1$ means $\sqrt{\dfrac{1}{\omega^2}-1} \ll 1 \Rightarrow \omega \approx 1$. With this in mind, the approximate solution, obtained by decomposing (3) into a series, takes the form

$$u = 4\sqrt{\frac{1}{\omega^2}-1}\,\frac{\cos\left(\gamma\omega t - \sqrt{\gamma^2-1}\omega x\right)}{\cosh\left(\gamma x\sqrt{1-\omega^2} - t\sqrt{1-\omega^2}\sqrt{\gamma^2-1}\right)} \quad (15)$$

and coincides with (14).

## 2. Example - an approximate breather solution of the equation for the vector potential of an electromagnetic field in a graphene superlattice

As an example, we will consider the equation describing the propagation of solitary electromagnetic waves in a graphene superlattice (GSL) [13-15]:

$$\frac{\partial^2 u}{\partial t^2} - c^2 \frac{\partial^2 u}{\partial x^2} + \frac{\omega_0^2 b^2 \sin u}{\sqrt{1+b^2(1-\cos u)}} = 0. \tag{16}$$

Here $u = edA_z/\hbar c$ is the dimensionless component of the vector potential in the direction of alternating layers of SL, $\omega_0^2 = \frac{2\pi n_0 e^2 d^2 \Delta}{a_0 \hbar^2}$, $n_0$ is the surface concentration of charge carriers, $a_0 = 0.12$ is the thickness of the graphene layer, $b = \Delta_1/\Delta$, and the parameters $\Delta$ and $\Delta_1$ can be conditionally called the half–widths of the forbidden and allowed minizones, respectively, $d$ is the period of SL. It is assumed that the alternation of layers is carried out along the z axis. Equation (12) has a solution in the form of $2\pi$-pulse, expressed implicitly [13]:

$$\int_\pi^{u(\xi)} \frac{du}{\sqrt{\sqrt{1+b^2(1-\cos u)}-1}} = 2\xi, \tag{17}$$

$\xi = \frac{x-vt}{L_0}$, $L_0 = \frac{c}{\omega_0}\sqrt{1-\frac{v^2}{c^2}}$, $v$ is the velocity of the electromagnetic pulse.

Transform the equation (16). Let's introduce new variables:

$$\frac{x\omega_0 b}{c} \to x, \quad t\omega_0 b \to t. \tag{18}$$

Equation (16) takes a form:

$$\frac{\partial^2 u}{\partial t^2} - \frac{\partial^2 u}{\partial x^2} + \frac{\sin u}{\sqrt{1+b^2(1-\cos u)}} = 0. \tag{19}$$

Expanding $\frac{\sin u}{\sqrt{1+b^2(1-\cos u)}}$ into a series on $u$ up to cubic terms, we get

$$\frac{\partial^2 u}{\partial t^2} - \frac{\partial^2 u}{\partial x^2} + u - \left(\frac{b^2}{4} + \frac{1}{6}\right)u^3 = 0, \qquad (20)$$

so

$$\beta = \frac{b^2}{4} + \frac{1}{6}. \qquad (21)$$

According to (11), (13), the solution of equation (19) in the form of a breather can be presented as:

$$u = \left(\frac{32(1-\omega^2)}{3b^2+2}\right)^{1/2} \frac{\cos\left(\gamma\omega t - \omega x\sqrt{\gamma^2-1}\right)}{\cosh\left(\gamma x\sqrt{1-\omega^2} - t\sqrt{1-\omega^2}\sqrt{\gamma^2-1}\right)}. \qquad (22)$$

Turning to the original notation, we get:

$$u = \left(\frac{32(1-\omega^2)}{3b^2+2}\right)^{1/2} \frac{\cos\left(t\frac{\gamma\omega}{\omega_0 b} - x\frac{\omega c}{\omega_0 b}\sqrt{\gamma^2-1}\right)}{\cosh\left(x\frac{\gamma c}{\omega_0 b}\sqrt{1-\omega^2} - \frac{t}{\omega_0 b}\sqrt{\gamma^2-1}\sqrt{1-\omega^2}\right)}. \qquad (23)$$

It is of interest the study of the stability of the approximate solution (22). Using the Wolfram Mathematica package, we perform a numerical solution of equation (19), taking the function (22) as the initial condition. Figure 1 shows graphs of approximate analytical and numerical solutions at various points in time. When plotting the graphs, it was assumed $b=0.90$ and $\omega=0.97$. From the graphs presented in Figure 1, it can be seen that, although the condition $|u| \ll 1$ is not met at these values, the numerical solution turns out to be close to the analytical one presented above and shows stability, that is, the scope of applicability of the approximate analytical solution turns out to be somewhat wider than originally assumed.

Figure 2 shows the dependence of the solution on time at a fixed $x$. From this figure it can be seen that the pulse duration is approximately 20 units along the time axis. Figure 3 shows a graph of the solution when time $t$ is postponed along one of the axes, and a spatial coordinate $x$ is postponed along the other axis.

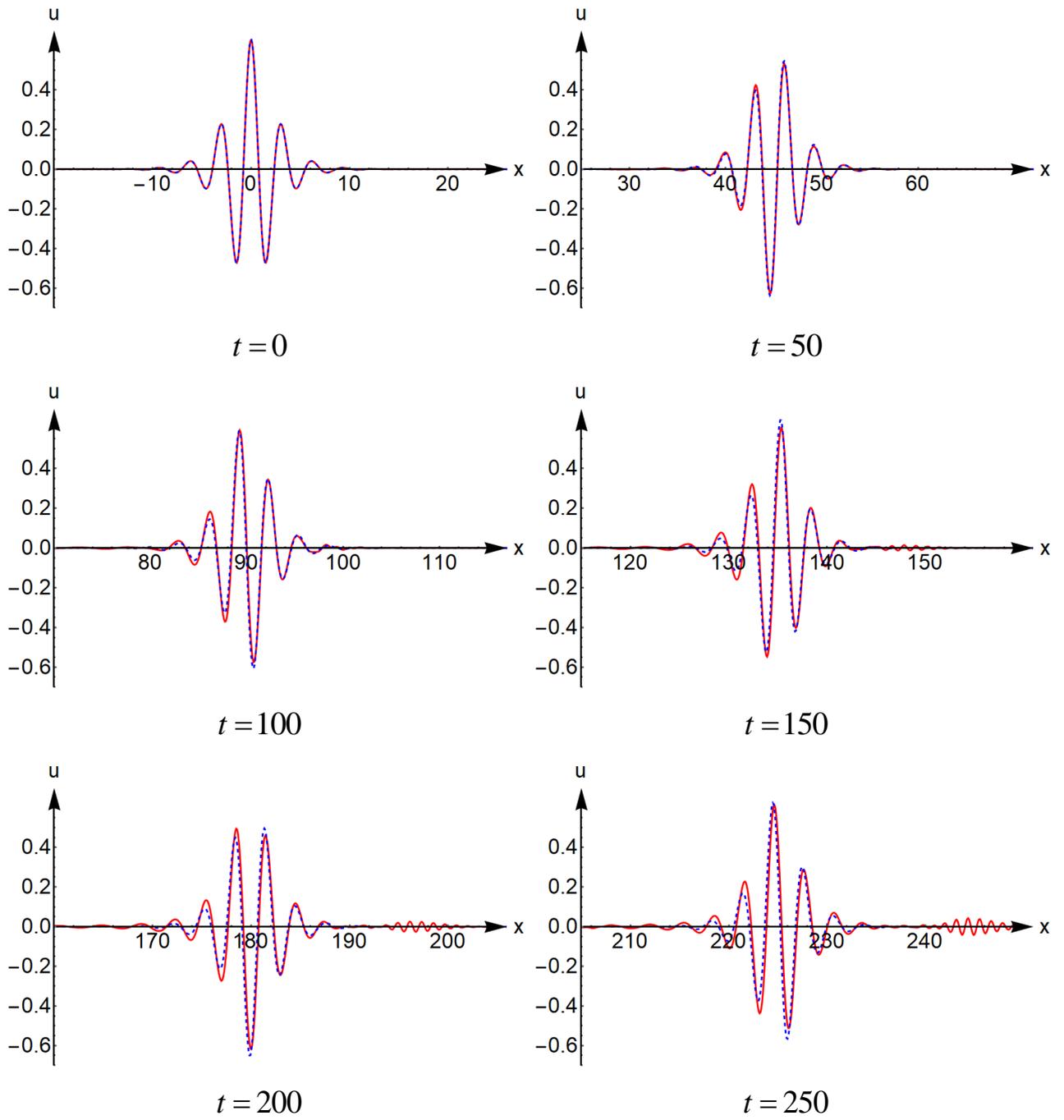

Figure 1 - Comparison of numerical (red line) and approximate analytical (blue dashed line) at different time points

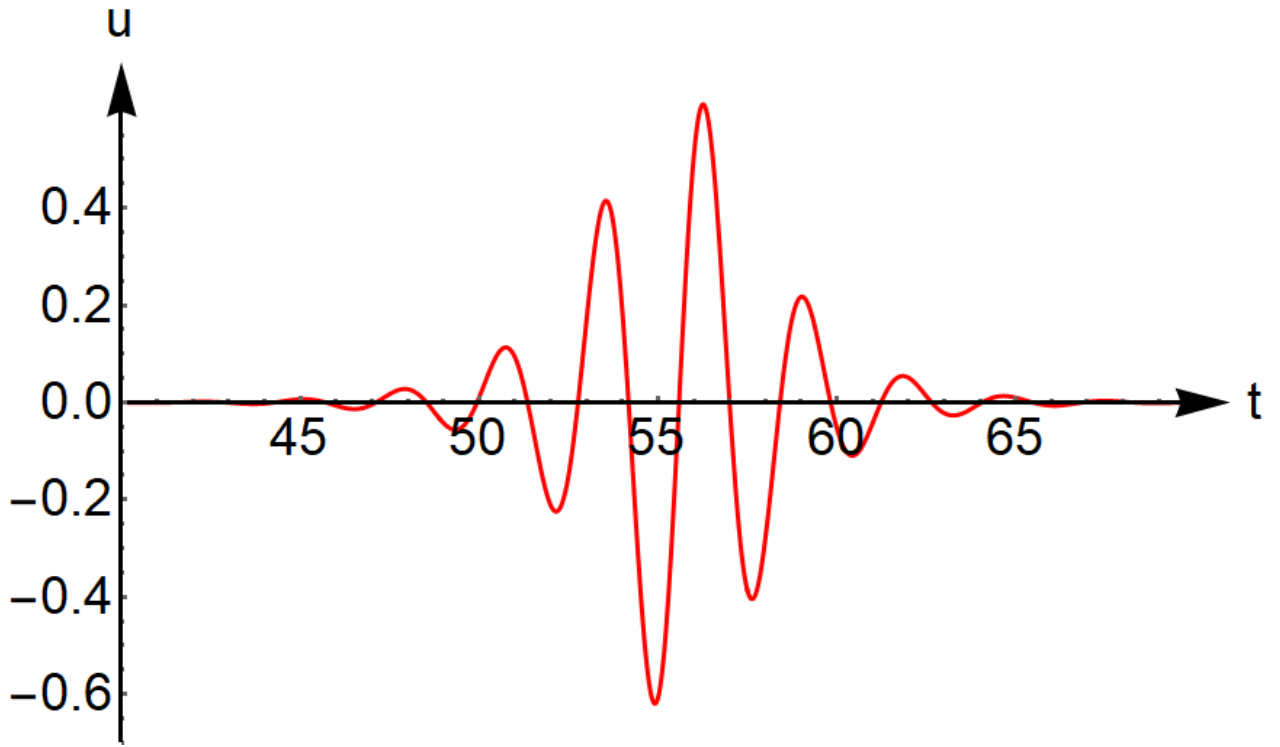

Figure 2 – Dependence of the approximate solution on time at $x = 50$

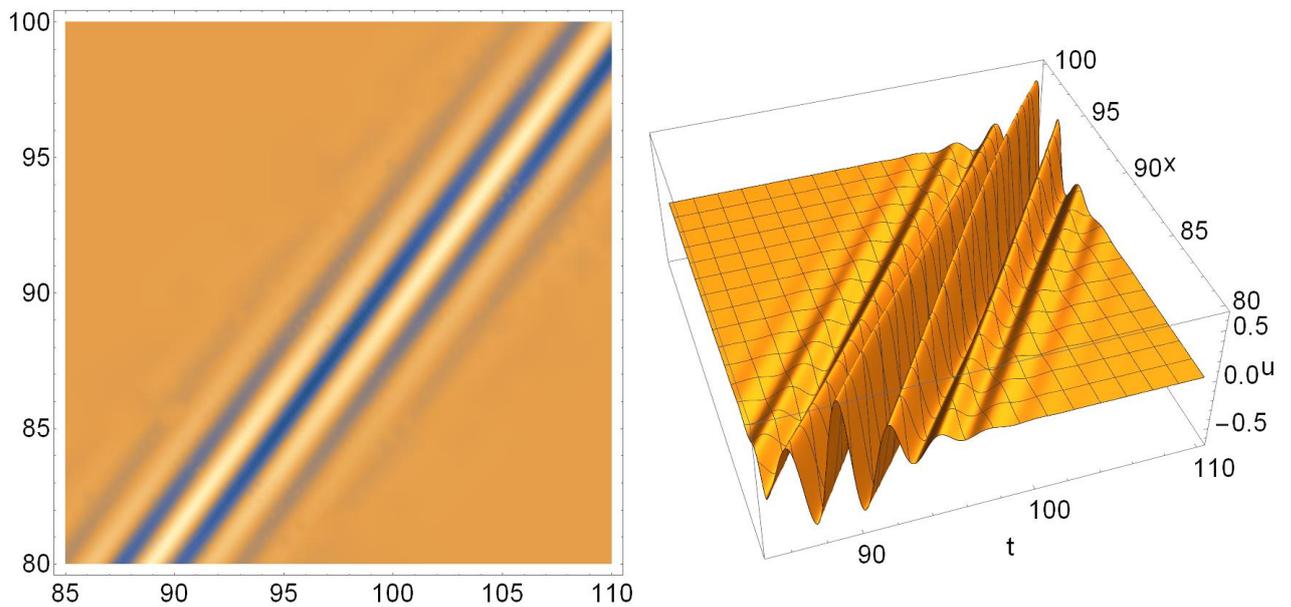

Figure 3 – General view of the approximate solution in the field
$t = \{85, 110\}$, $x = \{80, 100\}$

To quantify the differences between the numerical solution and the approximate analytical solution, we use the following technique. At given value of $t$ in the domain where the solution takes solutions other than zero (that is, $|u|$ is greater than some small positive value $\varepsilon$), we calculate the extremes of the numerical solution. Based on the list of values $\{x_i, |u(x_i,t)|\}$ where $x_i$ is determined the position of the extremes of the numerical solution, we construct an interpolation function. Figure 4 shows the graphs of the interpolation function which is the envelope of the absolute value of the numerical solution near the maximum amplitude at various points in time. Having found the position $x_{max}(t)$ of the maximum of the interpolation function, we determine the segment $[x_{max}(t) - L, x_{max}(t) + L]$, where $L$ is a half of the pulse width in space (judging by Figures 1, 4, $L \approx 10$). Next, we randomly select N values $\{x_i\}_{i=1..N}$ from this segment and form two vectors: $a = \{u_{appr}(x_i)\}_{i=1..N}$ and $b = \{u_{num}(x_i)\}_{i=1..N}$ which is the values of the approximate analytical and numerical solutions, respectively, at points $\{x_i\}_{i=1..N}$. To compare the approximate analytical and numerical solutions, we calculate the correlation coefficient at different points in time:

$$K_{corr} = \sum_{i=1}^{N} \frac{(a_i - \bar{a})(b_i - \bar{b})}{\sigma_a \sigma_b (N-1)}. \tag{24}$$

Here $\bar{a} = \dfrac{\sum_{i=1}^{N} a_i}{N}$ is the average sample value, $\sigma_a = \sqrt{\dfrac{\sum_{i=1}^{N}(a_i - \bar{a})^2}{N-1}}$ is the standard deviation. Figure 5 shows the dependence of the correlation coefficient between the numerical solution and the approximate analytical solution on time. It can be seen that the correlation coefficient decreases monotonically, however, even at the moment $t = 250$ $K_{corr} \approx 0.94$, which suggests that the proposed approximate solution in the form of a running breather decays quite slowly.

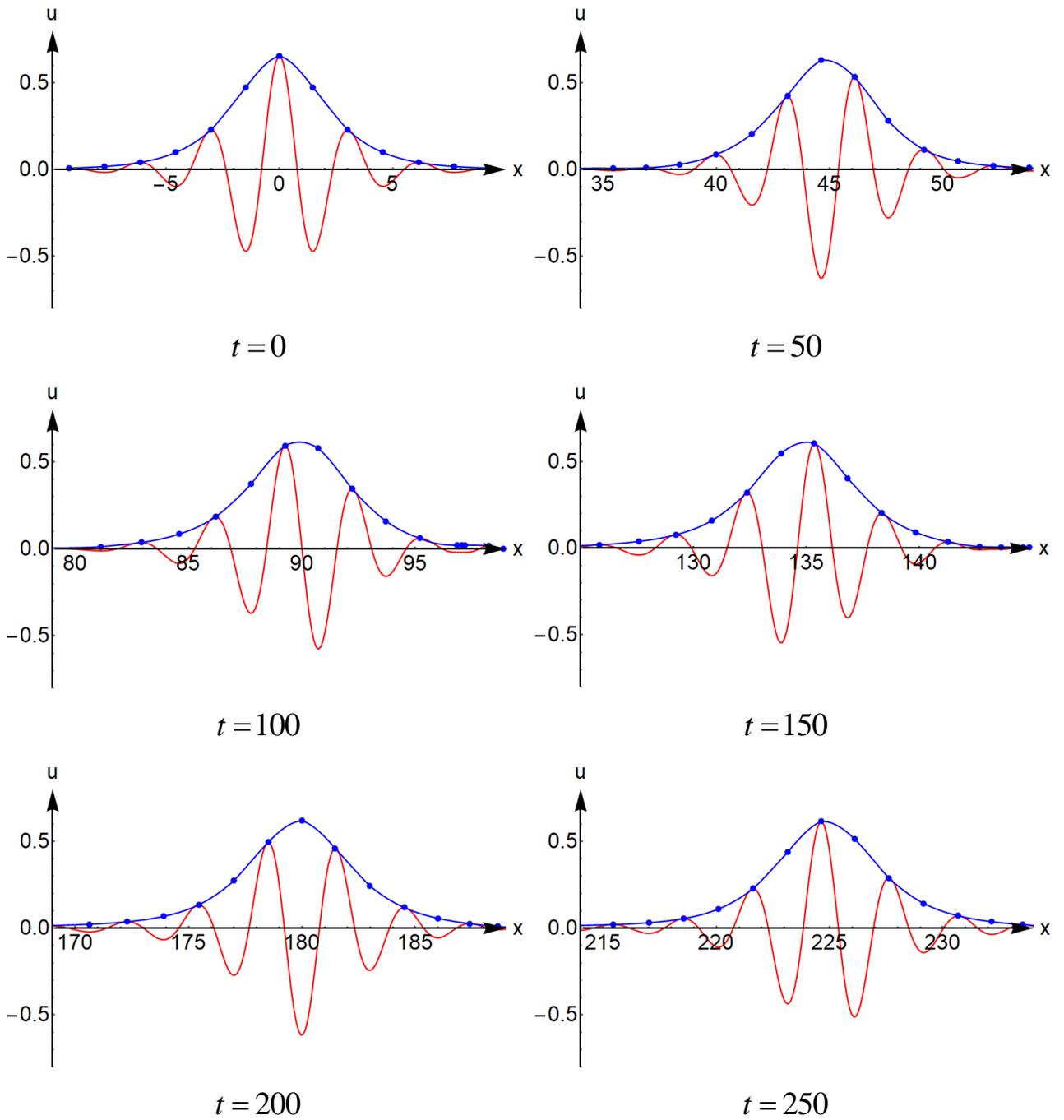

Figure 4 – Graphs of the interpolation function which is the envelope of the absolute value of the numerical solution (blue line and dots) near the maximum at various points in time

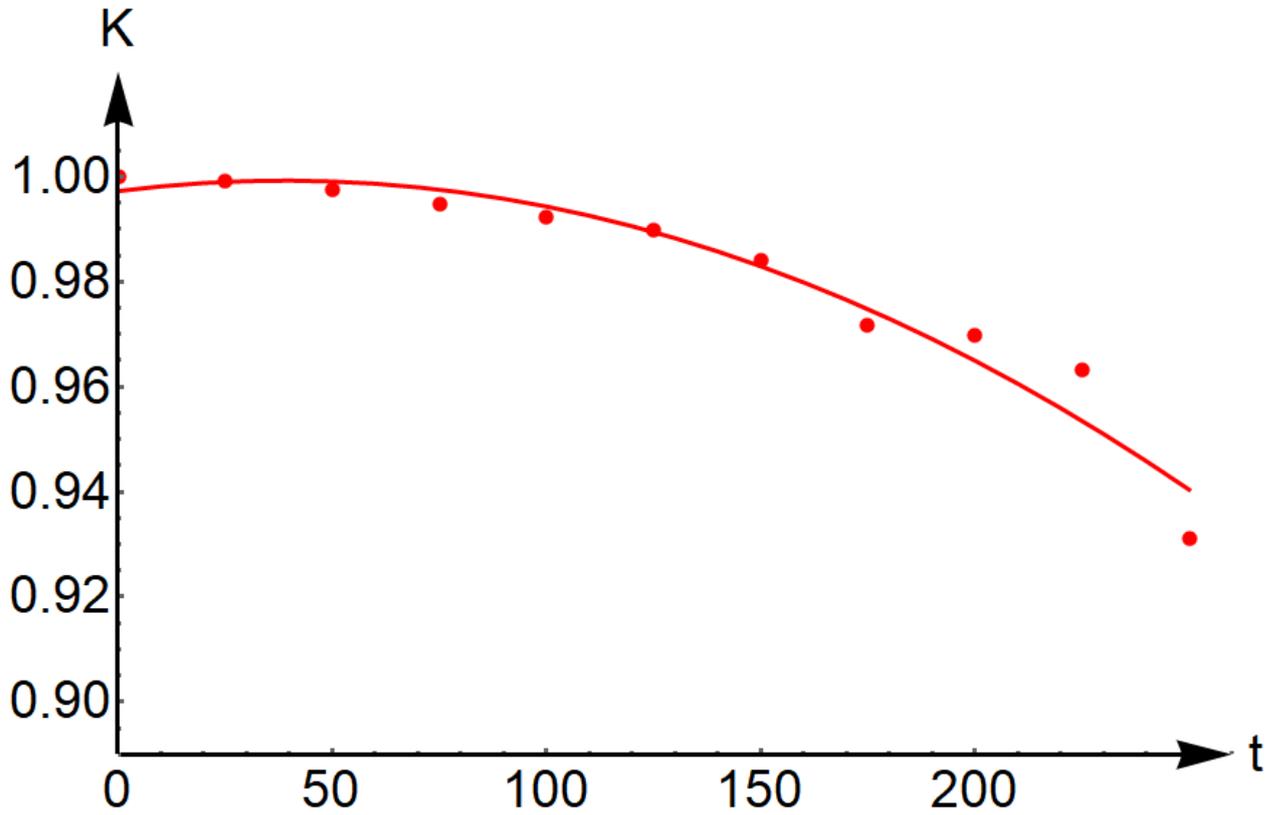

Figure 5 – Dependence of the correlation coefficient between the numerical and approximate analytical solution on time

## 3. Comparison with the works of other authors. Discussion of the results

In [16], it is considered a two-component breather solution of the nonlinear Klein-Gordon equation

$$\frac{\partial^2 U}{\partial t^2} - C\frac{\partial^2 U}{\partial x^2} = -\alpha_0^2 U + \frac{\alpha_0^2}{6}U^3, \qquad (25)$$

obtained by decomposing the Taylor series of the right side of the sine-Gordon equation at $U \ll 1$. In [16], the generalized perturbative reduction method developed in [17-19] was used. The solution obtained in [16] has the form:

$$U(x,t) = A\operatorname{sech}\left(\frac{t - x/V_0}{T}\right)\{\cos(k_1 x - \omega_1 t) + B\cos(k_2 x - \omega_2 t)\}. \qquad (26)$$

Solution (26) is similar in form to the solutions (14), (22) obtained in this paper. The essence of the perturbation reduction method used in [16-19] is the search for

a solution to a nonlinear partial differential equation in the form of an amplitude-modulated plane wave:

$$U = \sum_{\alpha=1}^{\infty} \varepsilon^{\alpha} U^{(\alpha)},$$

$$U^{(\alpha)} = \sum_{l=-\infty}^{+\infty} U_l^{(\alpha)}(\tau, \chi) \exp\left[il(kx - \omega t)\right]$$

(27)

where $\varepsilon$ is some small parameter, $\tau = \varepsilon^2 t$, $\chi = \varepsilon(x - \lambda t)$, $\lambda = \partial \omega / \partial k$ is a group velocity, $U_l^{(\alpha)} = U_{-l}^{(\alpha)*}$. Thus, the authors of [17], solving an equation of the form (25) for arbitrary values of the coefficients in front of the linear and cubic terms on the right side, obtain a nonlinear Schrodinger equation for the modulating function. However, as shown in [18], the solution of equation (25) taken in the form of a plane wave with a modulated amplitude turns out to be unstable. From our point of view, the method proposed in this paper for obtaining an approximate solution of the nonlinear Klein-Gordon equation in the form of a small-amplitude breather has an advantage over the perturbation reduction method used in [16-19] due to its simplicity.

The paper [15] is devoted to the study of breather solutions of equation (16) describing the propagation of nonlinear waves in a graphene superlattice. In [15] it is numerically studied the inelastic collision of kinks and antikinks, which are described by (17) and propagated with the same but opposite speed. After interaction of kink and antikink they escape to infinity when its speed is either larger than a critical value or it is inside a series of resonance windows. Otherwise, they form a breather-like state that slowly decays by radiating energy. The fractal structure of these resonance windows is characterized by using a multi-index notation and their main features are compared with the predictions of the resonant energy exchange theory showing good agreement.

Thus, the paper presents a method for obtaining an approximate solution of the nonlinear Klein-Gordon equation, which is a traveling breather of small amplitude. An example of obtaining such a solution for an equation describing the propagation of nonlinear waves in a graphene superlattice is considered. The analysis of the obtained solution for stability is carried out. It is shown that the shape of the solution changes weakly over a time interval of tens of pulse durations.